\documentclass[preprint,12pt]{elsarticle}

\usepackage{amssymb}

\journal{NIM A}

\begin{document}

\begin{frontmatter}



\title{Production of short lived radioactive beams of radium}
\author{P.D. Shidling\corref{cor1}}
\cortext[cor1]{Corresponding author}
\ead{P.Shidling@rug.nl}
\author{G.S. Giri, D.J. van der Hoek, K. Jungmann, W. Kruithof, C.J.G. Onderwater, M. Sohani, O.O. Versolato, L. Willmann,
\\
H.W. Wilschut}
\address{Kernfysisch Versneller Instituut, University of Groningen, Zernikelaan 25, 9747AA Groningen, The Netherlands}

\begin{abstract}
Short lived $^{212,213,214}$Ra isotopes have been produced at the 
TRI$\mu$P facility in inverse kinematics via the fusion-evaporation 
reaction $^{206}$Pb+$^{12}$C at 8 MeV/u. Isotopes are separated 
from other reaction products online using the TRI$\mu$P magnetic 
separator. The energetic radium (Ra) isotopes at the exit of the separator 
were converted into low energy ions with a thermal ionizer. 
Ra isotopes have been identified by observing their $\alpha$ 
decay and life times.
\end{abstract}

\end{frontmatter}


\section{Introduction}
In the TRI$\mu$P program \cite{ra1} radioactive beams are used to search 
for electric dipole moments (time reversal violation), atomic parity violation 
in heavy nuclei and to study the weak interaction via $\beta$ decay of 
light nuclei \cite{ra6,ra9}. The aim is to study physics beyond the standard model. 
The heaviest alkali-earth element is radium (Ra). Ra has unique atomic and 
nuclear properties which make it a very promising candidate for such 
experimental studies \cite{raf2,raf3}. Short life times require online production. 
Low-energy fusion-evaporation reactions can be used to produce Ra nuclei in the mass 
region A$<$215. These isotopes live long enough to be used in these studies. 
In particular A = 213 which has the ground state spin $\frac{1}{2}$ is of interest. 
For better understanding the details of the atomic spectroscopy of Ra, 
the study of an isotopic chain of Ra is relevant. The isotopes considered here 
are all $\alpha$ emitters: $^{212}$Ra (E$_{\alpha}$=6.9 MeV), $^{213}$Ra 
(E$_{\alpha}$=6.623 MeV, 6.713 MeV), and $^{214}$Ra (E$_{\alpha}$=7.137 MeV).
\par
There are two fundamental approaches with which one can produce different 
radioactive isotopes namely ISOL \cite{ra7} and in-flight separation \cite{ra8}. 
Here we use a combination of both. In this work Ra isotopes 
have been produced via low-energy fusion-evaporation reactions in inverse 
kinematic mode, thus using in-flight separation. The secondary beam is 
focused into a thermal ionizer and converted to a singly charged low energy 
ion beam, thus exploiting also ISOL techniques. The low energy is necessary 
to trap the ions. The advantage of this combined technique is the negligible 
activation of the thermal ionizer (TI). Different alkali and alkali-earth 
radioactive isotopes can be produced effectively using appropriate 
reactions such as fusion, fragmentation or direct reactions using the same TI, 
i.e. decoupling the production and thermalization process.

\section{Radium Production} 
The TRI$\mu$P dual magnetic separator \cite{ra13} consists of two dipole sections. 
The first is used to separate the products from the primary beam, 
whereas the second can provide further separation of reaction products and 
refocuses isotopes of interest at the final focal plane. Inverse kinematics
helps to produce secondary beams that match the angular and momentum 
acceptance of the separator. Secondary beams at the exit of the separator 
have a wide energy distribution and their energy is too high to use directly. 
Therefore, thermalization is required as mentioned before.
\par
For Ra production, a $^{206}$Pb beam of 8 MeV/u from the AGOR 
cyclotron bombarded a diamond-like-carbon (DLC) target \cite{radlc} of 
4 mg/cm$^{2}$ thickness. The projectile-target combination of Pb and C is chosen 
as it has a lower fission cross section compared to less asymmetric 
reactions. The isotope $^{206}$Pb has been chosen to maximize the 
cross section of $^{213}$Ra, while the target thickness assures that 
the excitation function for $^{213}$Ra production is fully used.
\par
The charge state distribution of the $^{206}$Pb beam after passing the target was determined by 
measuring the beam current after the first dipole section by varying the magnetic settings. 
In case of a thick diamond target, the beam charge states were not resolved. 
In order to better understand the charge state distribution a thin Al target of 
270 $\mu$g/cm$^{2}$ was used. Nine beam charge states were observed 
as shown in Fig.$\;$1. On the basis of these data and assuming that Ra has the 
same charge state distribution, we concluded that we can accept 36$\%$ of the 
charge state distribution with the momentum acceptance of 3$\%$ of the 
current configuration of the TRI$\mu$P magnetic separator \cite{radetail}.
\begin{figure}[h]
\begin{center}
\includegraphics[scale = 0.5, angle = 0]{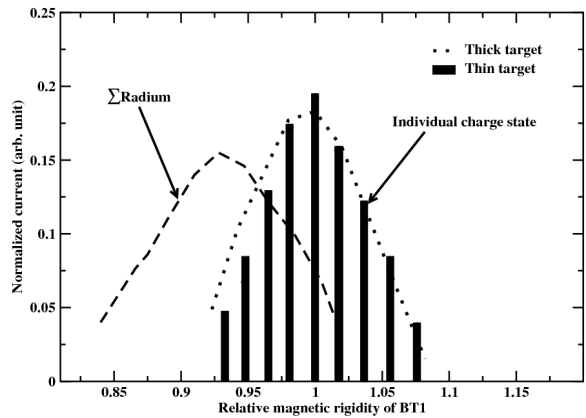}
\caption{\label {aa} Charge state distribution of the primary beam at intermediate focal plane 
after passing through a thick (dotted line) and a thin (Black bars) target. The dash line corresponds 
to the momentum distribution of radium products.}
\end{center}
\end{figure}
\par
At the exit of separator Ra production 
was confirmed by stopping the products of interest and the primary beam in 
a 80 $\mu$m thick Al catcher foil which was mounted in front of a silicon 
detector. Detector and foil were at an angle of 45$^{\circ}$ allowing the $\alpha$ 
particles from the decay of Ra to be detected, while stopping the primary beam. 
The magnet settings were later optimized as described below.
\par
The thermal ionizer (TI) \cite{ra14} 
consists of a stack of thin tungsten (W) foils of 1 $\mu$m thickness placed in a W cavity. The maximal 
W foil thickness of the cavity was estimated with the program SRIM \cite{ra15}. 
By heating the cavity to about T$\approx$2500 K the isotopes 
diffuse out of the foils and get ionized in collisions with the cavity 
walls. Extraction is done by applying a negative electric potential on 
the extraction electrodes of 6 to 10 kV. Ra isotopes released from the TI 
were stopped on a thin 1.8 $\mu$m thick Al foil in front of a silicon detector. 
The detector was calibrated with a composite $\alpha$ source ($^{239}$Pu,
$^{241}$Am and $^{244}$Cm). Only the $\alpha$ particles from the decay 
of Ra and its daughter nuclei were observed. Fig.$\;$2 
shows a typical $\alpha$-energy spectrum. $^{213}$Ra has two peaks with 
energies of 6.623 MeV and 6.713 MeV. These two peaks are not resolved due 
to the straggling in the Al foil in front of the silicon detector. 
Also for $^{212}$Ra an isolated peak is not seen. 
\begin{figure}[h]
\begin{center}
\includegraphics*[scale = 0.5, angle = 0]{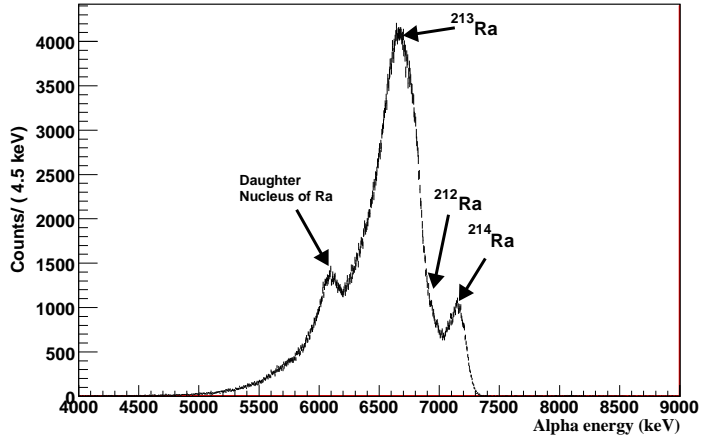}
\caption{\label {aa} Typical $\alpha$-energy spectrum observed after implanting Ra isotopes 
in a 1.8 $\mu$m Al foil in front of silicon detector. }
\end{center}
\end{figure}
\par
The separator magnet settings were optimized for Ra. 
The optimized setting for $^{213}$Ra was 8$\%$ lower than the setting for the 
primary beam and therefore the Ra was not fully separated from the primary 
beam due to the presence of some charge state of beam at this settings 
(see Fig.$\;$1). The distribution of Ra is broader because the various isotopes 
are produced with somewhat different average momenta. We note that the Pb beam 
after passing the DLC target has lost sufficient energy to minimize 
possible reactions in the W cavity. 
\par
In order to extract the production rate and the half life of the 
produced isotopes the TI output was studied in two 
different modes. In the first mode the extraction was continuously on 
and the primary beam was chopped in a cycle with 500$\;$s beam on and 
500$\;$s beam off. In the second mode the primary beam was 
continuously on and the extraction voltage of the TI was cycled 
with the same period. The chopping time of 500$\;$s was chosen to exceed the 
lifetimes of the Ra isotopes. In the extraction 
on/off mode the build-up and decay of the activity on the Al foil 
depends only on the lifetime of the extracted isotopes. In the primary 
beam on/off mode the apparent lifetime of the produced isotopes at the 
measuring site is influenced by the time the particle spends inside the TI. 
Therefore, in case of the beam on/off mode the output is delayed as compared 
to the extraction on/off mode. Measurements with beam on/off and 
extraction on/off mode were carried out at temperatures 2290 K, 
2400 K, 2430 K, 2480 K, and 2520 K, as measured with a pyrometer. Fig.$\;$3a and 3b shows 
the output of both modes at the lowest and highest temperature, 
respectively. At low temperatures, the output of the TI for beam 
on/off mode is delayed compared to that of the extraction on/off mode, 
whereas the outputs were nearly identical at higher temperature. We will use this 
temperature dependent information to characterize the TI in more 
detail. This will be reported elsewhere \cite{rashid}. The present result shows 
that the TI releases the Ra isotopes very fast at the highest temperature.
\begin{figure}[h]
\begin{center}
\includegraphics[scale = 0.45, angle = 0]{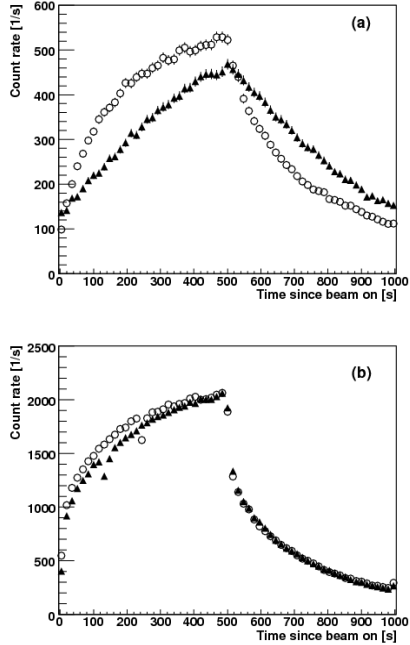}
\caption{\label {aa} Activity measured after the thermal ionizer for cyclotron 
beam switiching (solid triangle) and TI extraction switiching (open circle) 
at T = 2290 K (a) and T = 2520 K (b). The dips in the latter are associated with sudden drops in the beam intensity.}
\end{center}
\end{figure}
\newpage
\section{Data analysis}
The half life of the produced isotopes and their production rate was 
extracted by fitting the activity curve obtained from beam on/off 
and extraction on/off mode. The instantaneous dead time was measured 
by comparing the event rate in a scaler and the number of actual events processed 
by the data acquisition. The dead time was 30$\%$ to 40$\%$. To extract 
the life time and the production rate of $^{214}$Ra a gate was put on the 
$\alpha$ energy corresponding to $^{214}$Ra (Fig.$\;$2). Fig.$\;$4a shows the extracted 
build-up and decay activity curve for this gating condition. The activity curve 
was fitted by assuming a single characteristic life time (Fig.$\;$4a). 
Similarly the activity curves for $^{213}$Ra and $^{212}$Ra were 
extracted by gating on their respective $\alpha$ energies (shown in Fig.$\;$4b and 4c).
\begin{figure}[h]
\begin{center}
\includegraphics[scale = 0.5, angle = 0]{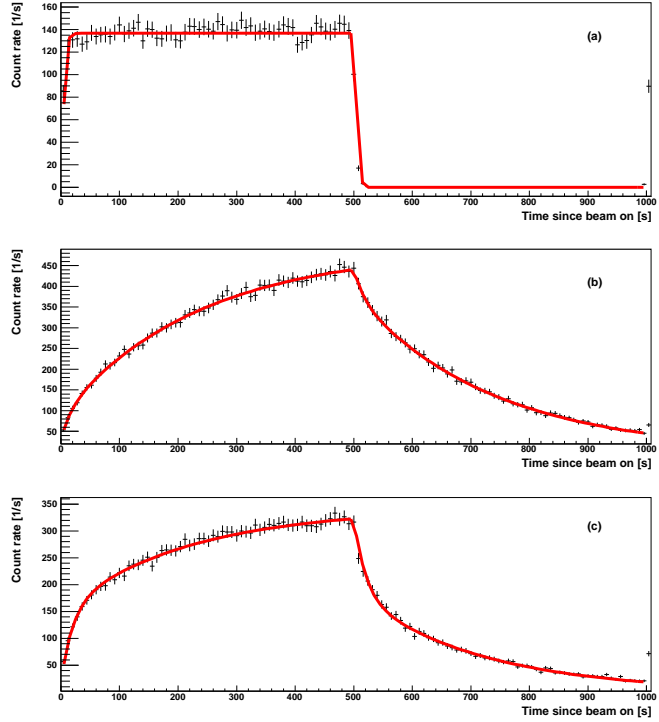}
\caption{\label {aa} Activity observed with a gate on the $\alpha$ energy region of $^{214}$Ra (a), 
$^{213}$Ra (b) and $^{212}$Ra (c) for TI extraction switching. The fit result is shown by line. 
The temperature of the TI is 2430 K.}
\end{center}
\end{figure} 
In case of $^{213}$Ra and $^{212}$Ra, the activity curve could not be fitted 
by a single characteristic life time because the $\alpha$-peaks of both isotopes 
were not resolved. Therefore the activity curve was fitted by 
considering two characteristic life times. The extracted characteristic lifetimes in 
both modes (beam on/off and extraction on/off mode) are tabulated in Table 1 
and compared with the literature values. 
\label{}
\begin{table}[h]
\begin{center}
\caption{\label{tab:table1}Measured life times for different Ra isotopes. 
The second column gives the actual life time from the extraction on/off mode, 
the third column refers to the switching of the beam and contains a delay time 
associated with the residing time in the TI. The last column gives the available 
literature value.}
\begin{tabular}{cccc}
\\
\\
\hline
\hline
\\
Isotope        &  Characteristic half lifes [s]      & Characteristic half lifes [s]   & Literature \\
              &   Extraction on/off mode            &     Beam on/off mode            &  Value \\     
\\
\hline
\hline
\\
$^{214}$Ra    &            2.42$\pm$0.14          &          3.48$\pm$0.12       &   2.46 $\pm$0.03 \cite{ra18} \\

\\
$^{213}$Ra    &            162$\pm$1.7            &          168$\pm$2.5         &   164.4$\pm$3.6  \cite{ra16}  \\
\\
$^{212}$Ra    &            12.5$\pm$1.0           &          15.8$\pm$1.0        &   13.0$\pm$0.2 \cite{ra19}\\
 
\\
\hline
\hline
\end{tabular}
\end{center}
\end{table}
Table 1 shows that there is a 
small difference in the characteristic life time between the two modes. 
This is consistent with our earlier conclusion that the TI 
releases the isotopes very fast, so that also the short-lived Ra isotopes 
can be extracted effectively. The characteristic life time from the 
extraction-on/off-mode activity curve (which depends exclusively on the life 
time of the isotopes) agrees with the literature value. In the present 
work the life time of $^{213}$Ra is more accurate as compared to the literature values. 
$^{213}$Ra and $^{214}$Ra, which are our particular interest, are produced respectively 
at a rate of 650$\;$s$^{-1}$ and 200$\;$s$^{-1}$ per particle nA of primary beam at the exit of the TI.
\newpage
\section{Conclusion and outlook}
In conclusion, $^{212,213,214}$Ra have been produced via low-energy 
fusion-evaporation reactions in inverse kinematics by bombarding a diamond target with a $^{206}$Pb beam 
of 8 MeV/u. Identification of different Ra isotopes 
has been done from the observed $\alpha$ spectrum and from the characteristic 
life time of the isotopes. Our technique involves a hybrid of in-flight 
production and ISOL methods and allows a flexible approach to the 
production of alkali and alkali-earth elements. Most actinides have a ionization 
potentials not much larger than Ra. Therefore, the production method 
used for Ra can be exploited for other actinides as well. For completeness 
we also note that we transported the produced isotopes successfully 
through a RFQ cooler and buncher \cite{ra1}. The current yields are more 
than sufficient to start our experimental program concerning parity violation 
and EDM searches. Production is limited by the maximum beam intensity that can be extracted 
from the AGOR cyclotron. A rotating target is currently being designed to 
be used with higher beam intensities.

\section{Acknowledgments}
This work was supported by the Stichting voor Fundamenteel
Onderzoek der Materie (FOM) under program 48
(TRI$\mu$P). We would like to thank the members of the AGOR cyclotron
group and the KVI support staff for valuable support.




\begin{thebibliography}{00}
\bibitem{radetail} The full momentum acceptance is $4\%$, which would improve the throughput to 45$\%$.
\end{thebibliography}
\end{document}